# Triclinic $Ni_{0.6}Co_{0.4}TiO_3$ Ilmenite Oxide

Yukari Fujioka,[1,a)] Johannes Frantti,[1,2] Anna Llobet,[3] Graham King,[3] and Steven N. Ehrlich[4]

[1]*Department of Applied Physics, Aalto University, 00076 Aalto, Finland*

[2]*Finnish Research and Engineering, Jaalaranta 9 B 42, 00180 Helsinki, Finland*

[3]*Lujan Neutron Scattering Center, Los Alamos National Laboratory, Los Alamos, New Mexico 87545, USA*

[4]*National Synchrotron Light Source, Brookhaven National Laboratory, Upton, New York 11973, USA*

Forming a solid-solution of $NiTiO_3$ and $CoTiO_3$, two isostructural (ilmenite) and isosymmetrical (space group $R\bar{3}$) compounds, result in a single-phase compound with a remarkably low crystal symmetry. By neutron and X-ray synchrotron powder diffraction techniques, the space group symmetry of the $Ni_{0.6}Co_{0.4}TiO_3$ sample was found to be triclinic $P\bar{1}$ at room temperature, far above the magnetic transition temperature. Ni and Co ions were found to prefer positions close to the octahedron center, whereas Ti ions took off-center positions. This structural distortion is the first known case in ilmenites and opens up ways to modify functional properties of magnetic oxides. Origin of the symmetry lowering is discussed.

## I. INTRODUCTION

Many ilmenite oxides $ATiO_3$, where $A$ is a divalent metal cation, are reported to possess a high-symmetry $R\bar{3}$ prototype structure with two formula units per rhombohedral primitive cell. Well-known examples are $NiTiO_3$, $CoTiO_3$, $FeTiO_3$ and $MgTiO_3$.[1-5] In this structure, the asymmetric unit contains $A$ cation at *2c*, Ti cation at *2c*, and O at *6f* Wyckoff position. The structure is layered; $AO_6$ and $TiO_6$ octahedra alternate along the hexagonal *c* axis. Each layer consists of two, slightly separated, two-dimensional triangular cation nets, termed double layer. While in layers the $AO_6$ and $TiO_6$ octahedra share edges, out-of-layers the octahedra are connected by sharing octahedral faces as shown in Fig. 1.

At low temperatures the transition metal ilmenite oxides $ATiO_3$ ($A$ = Mn, Fe, Co, Ni) exhibit various antiferromagnetic orderings, which frequently correspond to the doubling of the hexagonal *c* axis.[3] The magnetic transition can be accompanied by corresponding changes in the nuclei positions as allowed by the symmetry of the magnetically ordered phase, although a good approximation is to assume that the nuclei positions still correspond to the room-temperature $R\bar{3}$ phase.

The present work is dedicated to $Ni_{0.6}Co_{0.4}TiO_3$ (NCT), a solid solution of $NiTiO_3$ (NT) and $CoTiO_3$ (CT). According to literature, besides having identical crystal structures,[3,4]

---

[a)] Author to whom correspondence should be addressed. Electronic mail: yukari.fujioka@aalto.fi.

both NT and CT exhibit similar magnetic ordering, including the direction of the magnetic moments at and below 23[5] and 38 K,[4] respectively. One could expect that the solid-solution does not show any change in structure, however, the present work demonstrates that the NCT exhibits lower symmetry than its constituent members at room temperature, well above the magnetic transition point.

## II. EXPERIMENTAL

### A. Synthesis of $Ni_{0.6}Co_{0.4}TiO_3$

Polycrystalline stoichiometric Ni0.6Co0.4TiO3 samples were prepared through the conventional solid-state technique. NiO (99 %, Aldrich), CoO (≤ 99.99 %, Aldrich) and TiO2 (99 - 100.5 %, Riedel-de Haën) powders were mixed in the desired molar ratio, then pressed into a pellet and sintered in air at 1373 K. Phase quality was inspected by X-ray diffraction using CuKα radiation. Composition was checked by energy dispersive spectroscopy of X-rays employing JEOL JSM-7500FA analytical field-emission scanning electron microscope.

### B. Neutron powder diffraction experiments

Time-of-flight neutron powder diffraction (ToF-NPD) experiments were carried out at the Lujan Neutron Scattering Center using the high intensity powder diffractometer (HIPD). The powder sample was sealed in a vanadium can under He exchange gas atmosphere.

### C. Synchrotron measurements

X-ray diffraction data were collected at beamline X18A of the National Synchrotron Light Source at Brookhaven National Laboratory. A thin ceramic disk in transmission geometry was used to collect data. An area detector was used for counting diffracted intensity. The X-ray wavelength was 0.6198 Å. After centering, the diffraction rings were converted into the usual two-theta-intensity form by the Datasqueeze 3.0.0 program.[6]

### D. Rietveld refinement

Rietveld refinements were carried out using the General Structure Analysis System (GSAS).[7] NPD data collected on detector banks 5 (40°), 6 (-40°), 7 (14°) and 8 (-14°) and the X-ray diffraction data were used in the refinements comparing different structural models. For the best model also higher resolution data banks 1 (153°), 2 (-153°), 3 (90°) and 4 (-90°) were included. All nine histograms were used simultaneously to test the final structural model to obtain more reliable structural parameters. Simultaneous fit of X-ray and neutron powder diffraction data is challenging as there are always small differences in Bragg reflection positions. Another issue is that X-rays are less sensitive to oxygen positions and atomic displacement parameters. For this reason, we used an overall atomic displacement parameter. On the other hand, the resolution of X-rays is high, which was crucial for resolving possible peak splits or asymmetries.

**III. RESULTS**

**A. Structural models**

The synchrotron X-ray (SX) measurements clearly revealed peak splits and weak superlattice reflections at around 14.8 and 17.0° inconsistent with the prototype ilmenite structure as shown in Fig. 2. Many features are so weak that they could not be detected at all by a conventional laboratory XRD-device. Worth noting is that the prototype 222 and 444 reflections (003 and 006, respectively in terms of the hexagonal cell) are slightly shifted from the computed positions. This could indicate that there is a small structural modulation along the hexagonal $c$ axis. To model the additional features, lower symmetry candidates were searched by looking at the group-subgroup relationships.[8] The tested subgroups are listed in Table I. Triclinic cell is metrically close to the rhombohedral cell and the notations $1 \times 1 \times 1$ (Model #1) and $2 \times 2 \times 2$ (Model #5) indicate that the rhombohedral axes were approximately unchanged and doubled, respectively. Trigonal space groups were additionally tested and eventually rejected.

The two-theta range between 23 and 25° which shows two asymmetric peaks was notably informative. The asymmetry could not be fit by the Models #0 to #4, whereas the Model #5 gave a good fit, as is seen in Fig. 2 which shows a LeBail refinement based on the Model #5. Also the weak superlattice reflections are properly described by the Model #5. We assumed that Ni and Co occupy the same Wyckoff position in their statistical proportions.

The structural models were further tested by conducting Rietveld refinement for both eight neutron and one SX histograms simultaneously: Fits using Model #5 are given in Fig. 3. The model describes all reflections and intensities, even though the model "compromises" between different data banks and instruments. Metrically model #5 is close to the rhombohedral prototype structure in which all axes are doubled. The predicted model is consistent with the low-temperature magnetic structure reported for NT and CT, which is characterized by the doubling of the hexagonal $c$ axis. It is worth noting that magnetic ordering usually corresponds to a lower symmetry which puts less constraints on the atomic nuclei positions. Though new atomic displacements are allowed in the magnetically ordered phase, they are often hardly detectable. In contrast, here already at the room-temperature, paramagnetic primitive cell is large. Actually, no hexagonal $a$-axis doubling at low temperature was reported in the early neutron diffraction studies.[3,4] This could either be due to the lower resolution characteristic to the early neutron diffraction instruments, or it can be a characteristic feature of the solid-solution. Solubility gap could result in segregation of two ilmenite phases. However, the superlattice reflections, as shown in Figs. 2 and 3, indicate that the primitive cell must be significantly larger than the prototype cell.

**B. Testing cation mixing**

The presence of reflections between 7 and 10° in Fig. 2 and correspondingly between 4.0 and 4.8 Å in Fig. 3 indicates that the sample structure was closer to the ilmenite, rather than the corundum structure. Thus, the approximation of two types of cation-oxygen layers, one possessing $(Ni/Co)O_6$ octahedrons (termed Ni/Co-layer) and the other one possessing $TiO_6$ octahedrons (termed Ti-layer) is a feasible first approximation. Since the sizes of Ti, Ni and Co cations are quite similar, the possibility of mixed layers, Ti in a Ni/Co-layer and vice versa, was tested.

First, composition was locked and an overall atomic displacement parameter (ADP) was refined after which occupancies were refined. The compositions of all Ni/Co-layers were constrained to be the same, and similarly the composition of all Ti-layers was constrained to be the same. Further, the overall composition was constrained to the nominal value. Oxygen occupancy and the fraction of Ti in Ti-layer, constrained to be the same as (0.6Ni0.4Co) in Ni/Co-layer was close to unity, indicating that the interchange of cations is rather small at room temperature. During refinement the occupancies were slightly changed. A test with large, fixed ADP value yielded unphysical values for cation occupancies. The structural parameters corresponding to the refined overall ADP and subsequently refined occupancies are given in Table II.

To gain further understanding bond-valence sums (BVS)[9] were computed. The BVS method was applied to estimate the cation valences, based on the structure given in Table II. Table III summarizes the valences and oxygen octahedral volumes for each Ni/Co and Ti cation. The average valences of Ni/Co are slightly larger than the nominal valence, and Ti valences are slightly smaller than the nominal values. When cations were in their nominal positions, the average of the eight Ni/Co and the eight Ti valences were 1.86 and 4.04, respectively. The sum of the average cation valence, 5.90, is practically the expected nominal value. Thus, the structural model and parameters are reasonable.

Table III tabulates also the valences Ti has in Ni/Co-layers, and the valences Ni/Co have in Ti-layer. The average Ni/Co and Ti valences at interchanged positions were 2.61 and 2.88, respectively, which accords to the fact that the $TiO_6$ octahedrons are slightly smaller than the $(Ni/Co)O_6$ octahedrons. Refinements showed that oxygen occupancy and the fraction of Ti in Ti-layer, constrained to be the same as (0.6Ni0.4Co) in Ni/Co-layer, were close to unity. This indicates that no significant mixing occurs in room temperature. The mixing actually decreased the average sum of different cation valences, which suggests that either no mixing occurs at room temperature or the Ti position in Ni/Co-layer should be different from the position of the Ni/Co. Though the latter might be the case, testing such a model is very challenging as the number of structural parameters is very large. At elevated temperatures, mixing would be interesting possibility as it could result in structurally different phase(s). The high-temperature properties will be addressed by Raman and XRD techniques.

### C. Triclinic structure

The present data show that cations are displaced from the octahedron center, while shift of Ti is larger than Ni/Co. Similar structural features are seen in the prototype structure. Each $TiO_6$ octahedron has an electric dipole moment. The current structure possesses an inversion symmetry which results in the presence of layer with equal but opposite dipole moment. In the present solid-solution case, however, magnetic ions occupy positions closer to the center of octahedra while shifts of Ti cations become larger when compared to the prototype structure. This led to the smaller separation of two Ni/Co triangular nets along the hexagonal $c$ axis and larger separation of the two Ti triangular nets. The characteristic differences between the $P\bar{1}$ structure and rhombohedral structure are; (i) maximum distance between Ti and O became longer, being about 2.3 Å in two of three $TiO_6$ octahedra which share an edge of a common $TiO_6$ octahedron in basal plane (cf. Fig. 1 panel (b)), (ii) the shortest Ni/Co-Ni/Co distance in basal plane became shorter, being 2.8 Å while Ti-Ti distances were similar to the distances found in NT and CT, (iii) some of Ni/Co-Ti distances between octahedra sharing face (cf. Fig. 1 panel (c)) shortened to 2.7Å, which occurs in every second Ti-double layers. This is consisted with the observed position shift of 222 and 444 reflections from the triclinic model, see Fig. 2. In the prototype structure, each Ti layer is consisted of two flat triangular network of Ti ions separated along the $c$ axis direction, which together form a double layer. In the triclinic phase, the individual triangular nets are no longer flat.

### IV. DISCUSSION

The axes of the triclinic structure were obtained by doubling the $a$, $b$, and $c$-axes of the rhombohedral lattice. Such a lowering in translational symmetry can be linked to freezing of phonon mode(s) at a high symmetry point of the Brillouin zone (BZ) boundaries. We have tested three high symmetry points of the rhombohedral BZ, namely $(0,\frac{1}{2},0)$, $(\frac{1}{2},\frac{1}{2},0)$, and $(\frac{1}{2},\frac{1}{2},\frac{1}{2})$. The site symmetry of the two first points is $C_i$, while the third point has $C_{3i}$ symmetry and therefore among the vibrational modes at $(\frac{1}{2},\frac{1}{2},\frac{1}{2})$ are doubly degenerate modes. If the model is correct, new superlattice reflection(s) should emerge at $G^T_{h'k'l'} = G^R_{hkl} + q$, where $G^T_{h'k'l'}$ and $G^R_{hkl}$ are the reciprocal lattice vectors of the triclinic and rhombohedral phases, the wave vector $q = \frac{2\pi}{a_R}(\frac{1}{2},\frac{1}{2},\frac{1}{2})$, and $a_R$ is the lattice parameter of the rhombohedral cell. Indeed, the superlattice reflections $2\bar{2}2$ at 14.8° and 453 at 17.0° of the panel (a) in Fig. 2 fulfill the relationship with rhombohedral reflections 201 and 210 and 211, respectively.

Structural deformation described by freezing of low-frequency normal mode(s), or *soft mode(s)*, have been known in many metal-oxide compounds, perovskite oxides being a well-known example. To the best of our knowledge, similar symmetry lowering as reported here has not reported for ilmenite compounds with respect to temperature, pressure, cation substitution, or other variables. The cation masses of NCT are close to the corresponding cation masses of NT and CT and thus no significant changes in vibrational properties is expected by trivial consideration. The number of electrons is, however, different when both Ni and Co are divalent, as BVS result showed. This could be linked to the origin of the normal mode instability at the $(\frac{1}{2},\frac{1}{2},\frac{1}{2})$ BZ point. Besides Ti ion tends to shift from the center of oxygen octahedron. This tendency may also contribute to stabilize the structural distortion of NCT.

The present results suggest an interesting possibility to create ferroelectric and magnetic materials: Ti-layer being responsible for ferroelectricity, whereas the Ni/Co-layer being responsible for magnetism. Despite bulk ceramics possess inversion symmetry, the situation can be different in thin film in which biaxial strain can be used to lower the symmetry, as the case of SrTiO$_3$ shows[10]: Biaxial strain in epitaxial thin films favors ferroelectric over the non-ferroelectric state stable in bulk crystals.

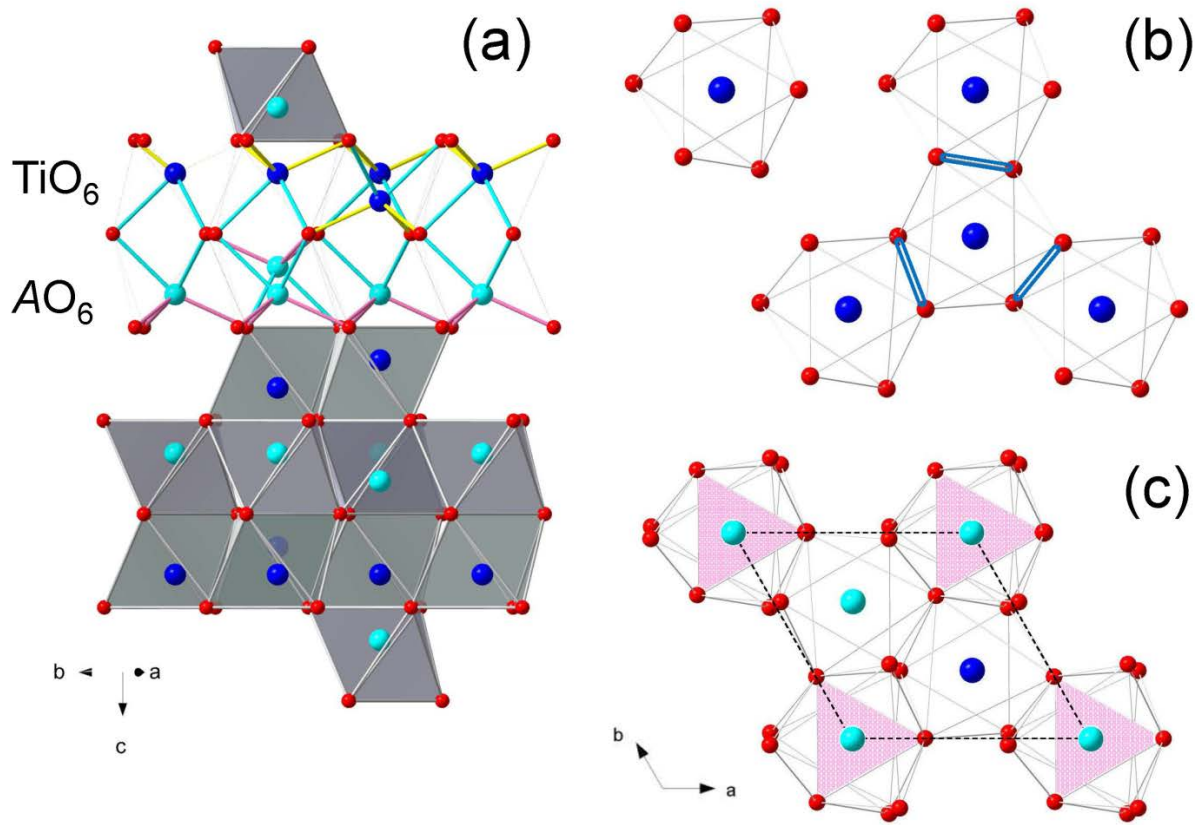

FIG. 1. The prototype $R\bar{3}$ ilmenite crystal structure $A$TiO$_3$ consists of alternating $A$O$_6$ and TiO$_6$ layers. The layers are perpendicular to the hexagonal $c$-axis and each layer consists of two triangular cation nets. The cation nets are slightly separated in the $c$ axis direction. Three shortest cation-oxygen bonds are indicated in yellow (Ti-O) and pink ($A$-O). Cyan indicates the three longest cation-oxygen bonds in an octahedron. Panel (b) shows the way oxygen octahedra edges are shared (blue double lines) within each layer, and panel (c) shows the way the oxygen octahedra of adjacent layers share faces (pink hatched).

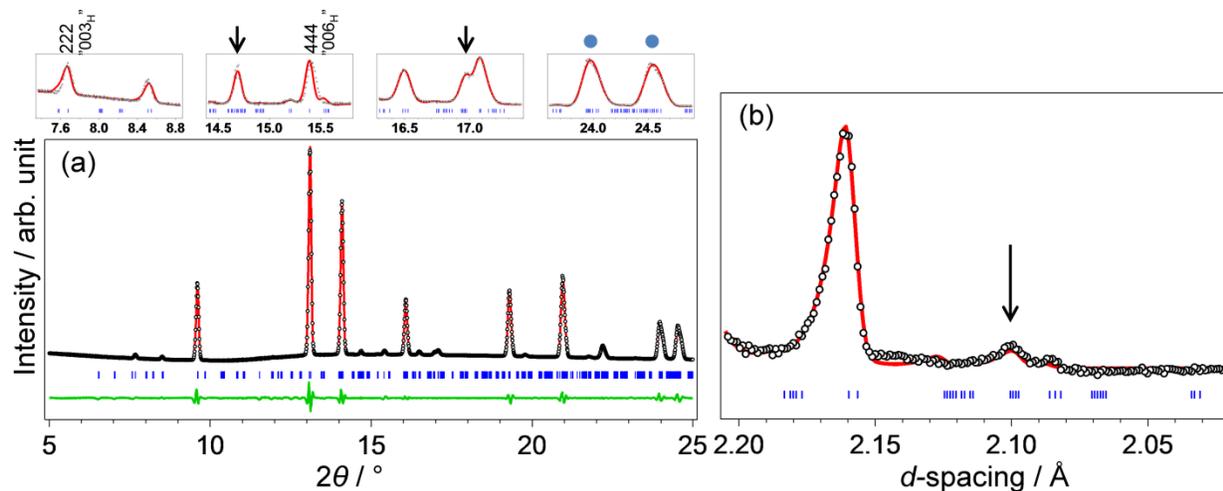

FIG. 2. LeBail refinement conducted for the (a) SX and (b) neutron diffraction data. The panel (a) shows the observed (black circles) and calculated (red line) intensities and the corresponding difference curve (green line) and Bragg reflection positions (blue tick marks) for the $P\bar{1}_{2\times2\times2}$ (#5) model. The upper left panels also reveal that the unambiguous features indicated by arrows at around 14.8° and 17.0° are modelled by the model #5. Evident is also the strong asymmetry of the peaks at 24.0 and 24.5 2θ° marked by circles, not explainable by the prototype phase. The panel (b) shows the superlattice reflection, indicated by arrows, NPD data corresponding to the range between 16.2 and 17.5° in panel (a).

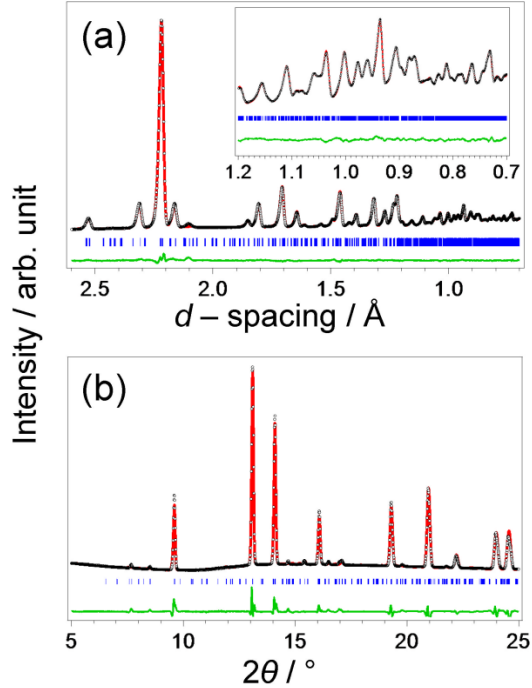

FIG. 3. Rietveld refinement of the neutron and SX data. Panel (a) shows the observed and calculated NPD intensities (153°) and the corresponding difference curves and Bragg reflection positions for the $P\bar{1}_{2\times2\times2}$ (#5) model. Panel (b) shows the corresponding curves for the SX data.

.

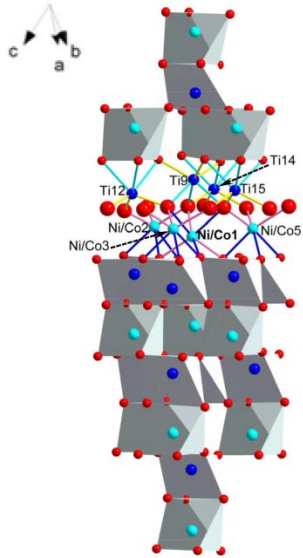

FIG. 4. Crystal structure of $(Ni_{0.6}Co_{0.4})TiO_3$ with $P\bar{1}$ symmetry. Cyan and blue lines in $TiO_6$ and $(Ni/Co)O_6$ layers represent three longest bond length in an octahedron and yellow and pink lines indicate three shortest bonds, respectively. Ni and Co ions shift to make six Ni/Co-O bonds roughly equal, while the displacement of Ti ions from the center of an octahedron becomes bigger. This is indicated by the Ni/Co1-O bonds where the one shortest and one longest bonds were swapped their orders.

TABLE I. Tested structural models and their relationship with the prototype phase $R\bar{3}$, Model #0. In the case of the $R\bar{3}$ symmetry three integers $l \times m \times n$ are used to indicate that the hexagonal $a_H$, $b_H$, and $c_H$ axes of the model are obtained from the prototype phase axes by multiplying them by $l$, $m$ and $n$, respectively. $\chi^2$ values were obtained by the refinements for NPD data ± 40° and ±14° and SX data using an isotropic atomic displacement parameter.

| No. | Relationship between prototype and model | $\chi^2$ |
|---|---|---|
| #0 | $R\bar{3}$ | 7.580 |
| #1 | $R\bar{3} \to P\bar{1}_{1\times1\times1}$ | 4.443 |
| #2 | $R\bar{3} \to R\bar{3}_{1\times1\times2}$ | 4.193 |
| #3 | $R\bar{3} \to R\bar{3}_{2\times2\times1}$ | 3.468 |
| #4 | $R\bar{3} \to R\bar{3}_{2\times2\times1} \to R\bar{3}_{2\times2\times2}$ | 8.595 |
| #5 | $R\bar{3} \to R\bar{3}_{2\times2\times1} \to R\bar{3}_{2\times2\times2} \to P\bar{1}_{2\times2\times2}$ | 2.753 |

TABLE II. Structural parameters, estimated from the Rietveld refinement conducted on HIPD neutron powder diffraction data and SX diffraction data of the $Ni_{0.6}Co_{0.4}TiO_3$ sample for $P\bar{1}$ symmetry and the statistical figures-of-merit (FOM). All atoms are at the Wyckoff position 2$i$. An overall ADP $U$ was refined. Since all data banks were used in this refinement, FOM are larger than those indicated in Table I.

| Lattice parameter: | $a$ / Å | 10.91013(10) | $\alpha$ / ° | | 55.0375(4) | |
|---|---|---|---|---|---|---|
| | $b$ / Å | 10.91436(10) | $\beta$ / ° | | 55.1176(4) | |
| | $c$ / Å | 10.92675(11) | $\gamma$ / ° | | 55.0517(4) | |
| $U$ / Å | | 0.005861(7) | | | | |
| Oxygen occupancy: | | 1.00636 | | | | |
| Mixing: | | 0.99931 | | | | |
| Atom | $x$ | $y$ | $z$ | Atom | $x$ | $y$ | $z$ |
| Ni/Co1 | 0.56814(24) | 0.59037(21) | 0.57202(24) | O17 | 0.36491(30) | 0.52965(33) | 0.72038(31) |
| Ni/Co2 | 0.57880(22) | 0.07772(22) | 0.56479(21) | O18 | 0.36041(29) | 0.05434(30) | 0.72608(29) |
| Ni/Co3 | 0.06808(22) | 0.57905(22) | 0.57959(22) | O19 | 0.86223(32) | 0.53668(33) | 0.72926(31) |
| Ni/Co4 | 0.07550(20) | 0.07289(21) | 0.56609(20) | O20 | 0.86828(30) | 0.03088(31) | 0.71558(32) |
| Ni/Co5 | 0.56852(23) | 0.57447(23) | 0.08427(22) | O21 | 0.36037(33) | 0.53056(34) | 0.22829(33) |
| Ni/Co6 | 0.56230(22) | 0.08302(23) | 0.07878(22) | O22 | 0.35733(32) | 0.03172(34) | 0.23725(30) |
| Ni/Co7 | 0.07405(19) | 0.57310(20) | 0.06409(19) | O23 | 0.86328(33) | 0.52997(34) | 0.22299(35) |
| Ni/Co8 | 0.07404(21) | 0.07225(20) | 0.07406(21) | O24 | 0.86483(30) | 0.03303(32) | 0.21894(32) |
| Ti9 | 0.67102(44) | 0.67753(41) | 0.67325(43) | O25 | 0.72129(31) | 0.36680(31) | 0.53722(32) |
| Ti10 | 0.67645(43) | 0.17856(41) | 0.67322(42) | O26 | 0.72965(32) | 0.86532(34) | 0.52819(35) |
| Ti11 | 0.17063(42) | 0.67753(41) | 0.68582(40) | O27 | 0.21873(34) | 0.36539(32) | 0.53383(34) |
| Ti12 | 0.16684(36) | 0.19147(34) | 0.69288(35) | O28 | 0.22806(30) | 0.85750(30) | 0.52716(30) |
| Ti13 | 0.67331(40) | 0.67478(39) | 0.18823(37) | O29 | 0.72267(31) | 0.35985(30) | 0.04333(34) |
| Ti14 | 0.67459(43) | 0.17966(42) | 0.17539(43) | O30 | 0.74994(32) | 0.84480(34) | 0.01597(32) |
| Ti15 | 0.17553(41) | 0.67763(40) | 0.17603(40) | O31 | 0.22975(33) | 0.35823(34) | 0.03330(36) |
| Ti16 | 0.18514(40) | 0.17600(40) | 0.17266(41) | O32 | 0.22961(31) | 0.86099(32) | 0.03967(34) |
| | | | | O33 | 0.51742(32) | 0.72302(33) | 0.37518(33) |
| | | | | O34 | 0.52594(33) | 0.21886(33) | 0.36811(33) |
| | | | | O35 | 0.02895(34) | 0.72585(30) | 0.36541(32) |
| | | | | O36 | 0.02285(34) | 0.21843(28) | 0.36662(28) |
| | | | | O37 | 0.53122(31) | 0.71180(30) | 0.87038(29) |
| | | | | O38 | 0.53369(33) | 0.21066(34) | 0.86942(32) |
| | | | | O39 | 0.02744(33) | 0.72113(31) | 0.87104(31) |
| | | | | O40 | 0.02928(35) | 0.22475(32) | 0.87249(33) |

| $\chi^2$ | 7.398 | | | | |
|---|---|---|---|---|---|
| Angle / ° | $R_{wp}$ / % | $R_p$ / % | Angle / ° | $R_{wp}$ / % | $R_p$ / % |
| +153 | 1.86 | 1.33 | −153 | 1.68 | 1.19 |
| +90 | 1.91 | 1.24 | −90 | 1.70 | 1.10 |
| +40 | 1.73 | 1.27 | −40 | 1.49 | 1.18 |
| +14 | 0.95 | 0.72 | −14 | 0.97 | 0.73 |
| SX | 8.91 | 6.27 | | | |

TABLE III. Bond-valence sums (BVS) of the cations listed in Table II in the cases of the cations occupied their nominal and interchanged positions. For Ni and Co ions, a composition weighed average is tabulated. Octahedron volumes are also given.

| Cation | BVS (nominal) | BVS (interchanged) | Octahedron volume / Å$^3$ |
|---|---|---|---|
| Ni/Co1 | 1.91 | 2.95 | 11.78 |
| Ni/Co2 | 2.00 | 3.09 | 11.20 |
| Ni/Co3 | 1.83 | 2.82 | 11.96 |
| Ni/Co4 | 1.82 | 2.81 | 11.98 |
| Ni/Co5 | 1.81 | 2.80 | 12.06 |
| Ni/Co6 | 1.72 | 2.65 | 12.58 |
| Ni/Co7 | 1.90 | 2.94 | 11.69 |
| Ni/Co8 | 1.92 | 2.96 | 11.51 |
| Ti9 | 3.90 | 2.53 | 10.10 |
| Ti10 | 4.24 | 2.75 | 9.56 |
| Ti11 | 4.08 | 2.64 | 9.86 |
| Ti12 | 4.14 | 2.68 | 10.12 |
| Ti13 | 3.84 | 2.48 | 10.01 |
| Ti14 | 3.93 | 2.54 | 10.38 |
| Ti15 | 3.76 | 2.43 | 10.41 |
| Ti16 | 4.41 | 2.86 | 9.34 |

## V. CONCLUSION

An ilmenite oxide solid solution, $Ni_{0.6}Co_{0.4}TiO_3$, made by mixing isostructural and isosymmetrical $NiTiO_3$ and $CoTiO_3$ was studied by neutron and synchrotron X-ray powder diffraction techniques. A mixture of Co and Ni at the same crystallographical site results in a triclinic $P\bar{1}$ symmetry. Metrically, the structure is close to the ideal rhombohedral structure, and approximately corresponds to the structure in which all axes of the prototype rhombohedral structure are doubled. New superlattice reflections of $P\bar{1}$ phase fulfilled the relationship $\boldsymbol{G}_{h'k'l'}^{T} = \boldsymbol{G}_{hkl}^{R} + \boldsymbol{q}$, which suggests that the structural distortion was originated by the freezing of doubly degenerate normal mode(s) at the $(\frac{1}{2},\frac{1}{2},\frac{1}{2})$ rhombohedral BZ point. Structural data analysis showed that the magnetic ions tend to occupy a position close to the octahedron center, whereas the Ti cations are shifted from the octahedron center. Bond-valence modeling showed that the Ni/Co and Ti cation valences were slightly larger and smaller than the nominal valence, their sum being very close to the expected nominal valence. The present structure has an inversion symmetry though, lower symmetry may be realized by introducing biaxial strain in thin film.

## ACKNOWLEDGMENTS

This work benefited from the use of HIPD at the Lujan Center at Los Alamos Neutron Science Center, funded by the U.S. Department of Energy (DoE), Office of Basic Energy Sciences (BES). Los Alamos National Laboratory is operated by

Los Alamos National Security LLC under DoE Contract DE-AC52-06NA25396. Use of the National Synchrotron Light Source, Brookhaven National Laboratory, was supported by the U.S. Department of Energy, Office of Science, Office of Basic Energy Sciences, under Contract No. DE-AC02-98CH10886. The research work was supported by the collaboration project between the Center of Excellence for Advanced Materials Research at King Abdulaziz University in Saudi Arabia (Project No. T-001/431) and the Aalto University.

**REFERENCES**

[1] Y. Fujioka, J. Frantti, R. M. Nieminen, J. Phys. Chem. C **115**, 1457 (2011).

[2] Y. Fujioka, J. Frantti, R. M. Nieminen, Mater. Sci. Forum **700**, 23 (2012).

[3] G. Shirane, S. J. Pickart, Y. Ishikawa, J. Phys. Soc. Jpn. **14**, 1352 (1959).

[4] R. E. Newnham, R. P. Santoro, J. H. Fang, Acta Cryst. **17**, 240 (1964).

[5] Y. Ishikawa, S.Akimoto, J. Phys. Soc. Jpn. **13**, 1298(1958).

[6] P. A.Heiney, Datasqueeze 3.0.0 program. http://www.datasqueezesoftware.com.

[7] A. C. Larson, R. B. Von Dreele, General Structure Analysis System (GSAS) Los Alamos National Laboratory Report (LAUR 86-784. New Mexico, 2004).

[8] *International Tables for Crystallography, Volume A*, Fifth Edition, Edited by T. Hahn. Dordrecht, The Netherlands. Springer 2005.

[9] D. I. Brown, *The Chemical Bond in Inorganic Chemistry: The Bond Valence Model.* (Oxford University Press, New York, 2009).

[10] M. D. Biegalski, E. Vlahos, G. Sheng, Y. L. Li, M. Bernhagen, P. Reiche, R. Uecker, S. K. Streiffer, L. Q. Chen, V. Gopalan, D. G. Shclom, S. Trolier-McKinstry, Phys. Rev. B **79**, 224117 (2009); M. D. Biegalaski, Y. Jia, D. G. Shclom, S. Trolier-McKinstry, S. K. Streiffer, V. Sherman, R. Uecker, P. Reiche, Appl. Phys. Lett. **88**, 192907 (2006).